\documentclass[11pt]{article}

\usepackage{amssymb, amsmath,amsthm, mathrsfs, natbib}
\usepackage{ulem}
\usepackage{multirow}
\usepackage{latexsym}
\usepackage{epsfig}
\usepackage{booktabs}
\usepackage{graphicx}
\usepackage{xcolor}
\usepackage{amsmath,amsfonts}
\usepackage{float}
\usepackage{subcaption,authblk}  
\usepackage[colorlinks=true, citecolor=blue]{hyperref}
\usepackage{lineno}

\usepackage{comment}

\usepackage[margin=1in]{geometry}
\usepackage{setspace}
\newtheorem{Proposition}{Proposition}

\newtheorem{Corollary}{Corollary}

\newtheorem{Remark}{Remark}

\newcommand{\be}{\begin{equation}}
\newcommand{\ee}{\end{equation}}
\newcommand{\beqa}{\begin{eqnarray*}}
\newcommand{\eeqa}{\end{eqnarray*}}
\newcommand{\beqn}{\begin{eqnarray}}
\newcommand{\eeqn}{\end{eqnarray}}
\newcommand{\baa}{\begin{array}}
\newcommand{\eaa}{\end{array}}
\newcommand{\bcc}{\begin{center}}
\newcommand{\ecc}{\end{center}}
\newcommand{\btab}{\begin{tabular}}
\newcommand{\etab}{\end{tabular}}

\newcommand{\mb}{\makebox}

\newcommand{\lb}{\label}

\newcommand{\nn}{\nonumber}

\newcommand{\mc}[1]{\mathcal{#1}}

\newcommand{\mcB}{\mathcal{B}}
\newcommand{\mcC}{\mathbb{C}}

\newcommand{\mcL}{\mathcal{L}}

\newcommand{\mcP}{\mathcal{P}}

\newcommand{\Ind}{1\!\mathrm{l}}

\newcommand{\tr}{{\rm tr}}

\newcommand{\diag}{{\rm diag}}

\newcommand{\argmin}{\mb{\rm argmin}}

\newcommand{\bsigma}{\boldsymbol{\sigma}}

\newcommand{\btheta}{\boldsymbol{\theta}}
\newcommand{\bTheta}{\boldsymbol{\Theta}}

\newcommand{\blambda}{\boldsymbol{\lambda}}
\newcommand{\beeta}{\boldsymbol{\eta}}
\newcommand{\bbeta}{\boldsymbol{\beta}}

\newcommand{\bdelta}{\boldsymbol{\delta}}

\newcommand{\bXi}{\boldsymbol{\Xi}}

\newcommand{\bSigma}{\boldsymbol{\Sigma}}

\newcommand{\bmu}{\boldsymbol{\mu}}

\newcommand{\bzeta}{\boldsymbol{\zeta}}

\newcommand{\bA}{\boldsymbol{A}}

\newcommand{\bB}{\boldsymbol{B}}

\newcommand{\bD}{\boldsymbol{D}}
\newcommand{\bdd}{\boldsymbol{d}}

\newcommand{\bee}{\boldsymbol{e}}

\newcommand{\bH}{\boldsymbol{H}}
\newcommand{\bI}{\boldsymbol{I}}

\newcommand{\bK}{\boldsymbol{K}}
\newcommand{\bk}{\boldsymbol{k}}

\newcommand{\bM}{\boldsymbol{M}}

\newcommand{\bP}{\boldsymbol{P}}

\newcommand{\bQ}{\boldsymbol{Q}}

\newcommand{\bR}{\boldsymbol{R}}
\newcommand{\br}{\boldsymbol{r}}

\newcommand{\bu}{\boldsymbol{u}}

\newcommand{\bv}{\boldsymbol{v}}
\newcommand{\bW}{\boldsymbol{W}}
\newcommand{\bw}{\boldsymbol{w}}
\newcommand{\bX}{\boldsymbol{X}}
\newcommand{\bx}{\boldsymbol{x}}
\newcommand{\by}{\boldsymbol{y}}
\newcommand{\bY}{\boldsymbol{Y}}

\newcommand{\bzero}{\boldsymbol{0}}
\newcommand{\bone}{\boldsymbol{1}}

\newcommand{\RR}{{\mathbb R}}

\def \nn{\nonumber}

\title{ Benchmarking Priors: A Fully Bayesian Approach to Small Area Benchmarking }
\author[1]{Soumadeep Bhowmick}
\author[1,2]{Anindya Roy}
\author[3]{Ryan Janicki}
\author[2,4]{Gauri S. Datta}

\affil[1]{University of Maryland Baltimore County, Baltimore, MD, USA}
\affil[3]{US Census Bureau, Washington, D. C. USA}
\affil[3]{US FDA, White Oaks, MD USA}
\affil[4]{University of Georgia, Athens, GA USA}

\usepackage{titling}
\pretitle{\begin{center}\LARGE\bfseries}
\posttitle{\par\end{center}\vspace{-0.6em}}

\preauthor{\begin{center}\large}
\postauthor{\par\end{center}\vspace{-0.2em}}
\date{}
\begin{document}

\maketitle
\vspace{-0.6in}
\begin{abstract}
We develop a new class of priors for Bayesian small area models under strict external benchmarking constraints. The prior is fully supported on the constrained parameter set defined by the benchmarking constraints and thereby provides a proper constrained posterior that is fully supported on the restricted set. The posterior naturally constrains the small area estimates to conform to the benchmarking values at the aggregate level. In addition, it facilitates valid uncertainty quantification using credible sets that are fully contained within the constrained set. 
This novel framework has a wide range of applicability, ranging from the standard Gaussian Fay-Herriot model to more general Fay-Herriot models, such as robust Fay-Herriot models,  where the area-level random effects are distributed as multivariate \(t\), or Fay-Herriot models where there is a complex nonlinear relationship between the area means and the covariates. We illustrate the usefulness of the proposed methodology using simulation experiments as well as by estimating the state-level proportion of  children  aged 5-17  living under poverty  based on the direct estimates  obtained from the Small Area Income and
Poverty Estimates (SAIPE) program of the U.S. Census Bureau and the national benchmark of the child poverty rate obtained from the  Annual Social and Economic Supplement of the Current Population Survey (CPS ASEC) conducted by the U.S. Census Bureau. The numerical results reaffirm the advantages of the proposed benchmarking priors.
\end{abstract}

\newpage

\section{Introduction}

Standard small area estimation (SAE) methods produce stable model-based estimates of various socio-economic or demographic variables for geographical areas with comparatively smaller sample sizes by borrowing strength from other areas.
These model-based estimates are often required  to conform to direct estimates
obtained at a higher level of aggregation. For example, when estimating population totals, model-based estimates for lower-level areas, such as counties, should, upon aggregation to the state level, equal the direct estimates for the state obtained from a larger survey. Although estimates at higher levels of aggregation could themselves be subject to error, they are often considered ``gold standard'' by policymakers. In the present article, we assume the external benchmarks obtained from a census or larger survey have no or negligible errors. If the direct estimates at the aggregated level are obtained from the census, the values may be constitutionally mandated to remain unchanged; hence, any aggregation method that sums up to the state-level total must produce a sum equal to the state total. Thus, small area estimates must be benchmarked to meet the constraints. The benchmarked estimates are expected to be more stable and resistant to misspecification of small area models.  

The problem of benchmarking  originated in the time series literature, where December values of monthly flow time series of inventory  had to be benchmarked to yearly time series of aggregate inventory; \cite{hillmer1987benchmarking, durbin1997benchmarking}.
In small area estimation, the problem of benchmarking also arises naturally and has been studied extensively from both frequentist and Bayesian perspectives. From a frequentist perspective, benchmarked estimators are generally obtained as constrained risk minimizers. Some of these approaches employ post-hoc adjustments of the unconstrained small area estimates to meet benchmarks (\cite{berg2009spree,fabrizi2012constrained}). In contrast, others incorporate benchmarks as constraints and then perform constrained estimation to obtain the small area estimates (\cite{pfeffermann2014single,pfeffermann2006small}). A third category involves embedding the benchmarking constraints in the estimation procedure, so that the estimators automatically satisfy the benchmarking constraints (\cite{wang2008small,bell2013benchmarking}). On the other hand, Bayesian approaches often involve minimizing posterior risk under benchmarking constraints (\cite{datta2011bayesian}, \cite{ghosh2013two}). However, most existing methods focus on point estimators and use ad hoc uncertainty measures.

The problem of benchmarking in the fully Bayesian context would mean that the aggregate of the small area means under the joint posterior of the individual means would have a posterior that is degenerate at  the specified value of the benchmark. In other words, any sample drawn from the posterior would automatically satisfy benchmarking constraints. This is particularly helpful in that it facilitates valid quantification of uncertainty for all small-area parameters while operating within the constrained parameter space. Recently, some studies (\cite{zhang2020fully}, \cite{okonek2024computationally}, \cite{sugasawa2024bayesian}, \cite{janicki2017benchmarking}) have adopted fully Bayesian approaches to enforce benchmarking. The methods of \cite{zhang2020fully} and \cite{okonek2024computationally} project unconstrained posterior samples onto the constraint space using a linear adjustment in the M-H step during posterior sampling. On the other hand, \cite{sugasawa2024bayesian} employs entropic tilting to identify the approximate posterior distribution whose expectation satisfies the benchmarking constraints, utilizing the reweighted unconstrained posterior samples to draw from the approximate posterior. While these methods provide benchmarked estimators, they rely on either ad-hoc adjustments or reweighting unconstrained posterior samples, lacking a formulation that inherently generates a posterior supported on the benchmarking constraints. 

In this article, we present a novel method that utilizes prior construction and reparameterization, which is fully Bayesian in the sense that it generates a valid proper posterior and the posterior samples automatically satisfy the benchmarking constraints. The key idea of our approach is to specify a class of priors that meet benchmarking constraints with probability one. Then, due to the Bayesian calculus, the posterior would satisfy the constraints with probability one. We refer to any such prior as \textit{a benchmarking prior} and the resulting posterior as the \textit{benchmarking posterior}. We apply our method to the basic area-level Fay-Herriot (FH) model (\cite{fay1979estimates}) and its various extensions. Building priors that are fully supported on the constrained set is an efficient way of performing full Bayesian inference under hard constraints, particularly if the constraints are complex enough to not support standard priors; e.g. \cite{gunn2005transformation, danaher2012minkowski,  roy2019constrained}. Our proposed method can accommodate multiple benchmarks and admits closed-form full conditionals, enabling straightforward implementation using Gibbs sampling.
\section{Benchmarking Prior}
\subsection{Fay-Herriot Model and Benchmarking}

Let \(\boldsymbol{Y} = \left( Y_1, \dots, Y_m \right)^\top\) be a vector of direct estimates obtained from sample survey data, typically using Horvitz-Thompson \citep{horvitz1952generalization} or H{\'a}jek-type estimators \citep{hajek1971comment}, which are usually design-unbiased estimators of the true small area means, \(\theta_i\).  It is often the case that there is a limited sample in one or more of the small areas, making some of the direct estimates, \(Y_i\), unreliable.  Precision can be improved by introducing a small-area model that leverages strengths across areas by incorporating auxiliary information.  The basic area-level model \citep{rao2015small} is given by
\begin{equation}\label{eq:Basic_Fay_Herriot}
  \begin{split}
   \bY & = \btheta + \bee, \\
   \btheta & = \bX \bbeta + \bu,
  \end{split}
\end{equation}
where \(\bee\) is a mean-zero random variable with known covariance, \(\boldsymbol{\Sigma}_e\), representing the sampling error, and \(\bu\) is a mean-zero random variable with covariance \(\boldsymbol{\Sigma}_u\), independent of \(\bee\).  The rows of the matrix \(\boldsymbol{X}\) are area-level covariates, \(\boldsymbol{x}^\top_i = (x_{i1}, \ldots, x_{ip})\), and \(\boldsymbol{\beta}\) is an unknown vector of coefficients.  The components of the model error, \(\boldsymbol{u}^\top = (u_1, \dots, u_m)\), are typically assumed to be i.i.d. mean zero Gaussian random variables with unknown variance \(\sigma^2_u\).  For illustration purposes, throughout this paper, we let \(\bee \sim N \left( \boldsymbol{0}, \boldsymbol{\Sigma}_e \right)\), but emphasize that the results are applicable with any mean zero sampling distribution with known covariance \(\boldsymbol{\Sigma}_e\) (although the computational details will be more involved in non-Gaussian settings).


In many applications, external estimates are available at a higher level of aggregation of small areas, which are often more reliable estimates of the underlying true features (such as means, totals, etc.) of the larger aggregate areas. The independent sources providing these higher-level values are treated as external as they are often from censuses or large-scale surveys, and hence the benchmarks   have no or negligible error allowing for the possibility of imposing the independently obtained values for the larger areas as {\it hard constraints} for the area-level means. This is the set up of the present article. These constraints can be linear or non-linear, with linear constraints being more common. Although the central idea of the proposed methodology is independent of the nature of the constraint, for the present article, we only consider linear constraints. Specifically, the area mean vector $\btheta$ is assumed to satisfy
\be
\bW^\top\btheta = \hat{\btheta}_{+},
\label{eq:constraints}
\ee
where $\hat{\btheta}_{+} = (\hat{\btheta}_{+,1}, \ldots, \hat{\btheta}_{+,q})^\top$ is a $q $ dimensional vector of external estimates of features of the larger aggregated areas. To avoid redundancy, we assume that $q < m$ and $\bW$ is an $(m\times q)$ full column rank matrix of weights used to construct the $q$ aggregates. Often $q$ will be much smaller than $m$.  Moreover, if $dim(\mcC(\bX) \bigcap  \mcC(\bW))  \ne 0,$  then there will be nontrivial vectors $\bv$ in $\mcC(W^\top)$ that will be internally benchmarked in the sense  $\bv^\top {\hat{\btheta}} = \bv^{\top}\by,$ where $\hat{\btheta}$ is some model-based estimate of $\btheta.$ This creates a conflict with the external benchmarking constraints. To avoid such situations, we  make the following assumption:\\

{\bf Assumption~1: [Full Rank]}  $\quad\quad \mcC(\bX) \bigcap  \mcC(\bW)   =   \{ \bzero \}. $\\

This assumption rules out redundancy between fixed-effect covariates and the benchmark constraints.  Related full-rank conditions arise in augmented-model approaches to self-benchmarking where the condition is given in terms of $\mcC(\bSigma_e \bW)$; see \cite{wang2008small} and \cite{bell2013benchmarking}. In \cite{bell2013benchmarking}, authors discuss the failure of a full rank condition when the benchmark is the simple aggregate and $\bX$ contains an intercept.

While the true area means are assumed to satisfy the constraints in \eqref{eq:constraints}, there is no guarantee that the estimated vector of means, ${\hat{\boldsymbol{\theta}}}$, will honor the constraints. In fact, the direct estimates of the small areas, $\bY$, may not satisfy the constraints, and any model-based estimator based on $\bY$ need not satisfy \eqref{eq:constraints} either. Requiring that the vector of estimated small area means, ${\hat{\boldsymbol{\theta}}}$, satisfies the benchmarking constraint \eqref{eq:constraints} often improves the efficiency of the estimator ${\hat{\boldsymbol{\theta}}}$ (cf. Corollary 2, (\cite{bell2013benchmarking})). This is the process of {\it external benchmarking}. Hereafter, we will refer to the constraints in \eqref{eq:constraints} as {\it benchmarking constraints}. Since we use a Bayesian approach to estimate the mean vector $\btheta$ based on the model \eqref{eq:Basic_Fay_Herriot}, we require that the posterior estimates of area-level means satisfy the aggregation constraints \eqref{eq:constraints}, and thus be \textit{benchmarked} to the external estimates at the higher level. 

\subsection{Constrained Set and Benchmarking Prior}
A Bayesian solution to the estimation of small-area means $\btheta$ can be formulated naturally in a Hierarchical Bayes (HB) formulation of the model \eqref{eq:Basic_Fay_Herriot}.  The HB formulation gives
\beqn
    \bY\mid\btheta,\bbeta,\bSigma_u\sim N(\btheta,\bSigma_e),\quad 
\btheta\mid\bbeta,\bSigma_u\sim N(\bX\bbeta,\bSigma_u),\quad 
(\bbeta,\bSigma_u)\sim\pi(\bbeta,\bSigma_u),
\label{eq:HB}
\eeqn 
where $\pi(\bbeta,\bSigma_u)$ denotes the joint prior distribution of $(\bbeta,\bSigma_u)$.  In standard setting, one assumes  $\bSigma_u = \sigma_u^2 \boldsymbol{I}$, where \(\boldsymbol{I}\) is the identity matrix, and the following prior specifications for $\bbeta$ and $\sigma_u^2$:
\[
\pi(\bbeta) \propto 1 \quad \text{and}\quad \sigma_u^2 \sim IG(a,b),
\] where $\text{IG}(\alpha,\beta)$ denotes an inverse-gamma distribution with shape $\alpha$ and rate $\beta.$ 
The above formulation has a closed-form full conditional distribution for each parameter, allowing a Gibbs sampler to conveniently draw samples from the joint posterior distribution of $(\btheta,\bbeta,\sigma_u^2)$. 
The Bayesian estimator ${E(\btheta|\bY)}$, as well as the posterior samples, under this standard formulation, do not necessarily satisfy the benchmarking constraints \eqref{eq:constraints}. 

We propose a novel framework to ensure that the posterior mean $E(\btheta | \bY)$ naturally satisfies the benchmarking constraints, and uncertainty quantification can be done based on the posterior samples that conform to the constraints. Under the benchmarking constraints on the area-level means, the parameter space is the affine linear subspace of $\RR^m$ given by
\be
\bTheta = \{ \btheta \in \RR^m:  \bW^\top\btheta = {\hat{\btheta}}_+ \}.
\label{eq:parameter_space}
\ee
One way of guaranteeing that the posterior mean $E(\btheta | \bY)$ belongs to the constrained parameter space $ \bTheta$ is to make sure the posterior is fully supported on this lower-dimensional affine subspace. Since the posterior support is a subset of the prior support, we can ensure that the posterior support is contained in $\bTheta$ by requiring the prior to be fully supported on $\bTheta.$ Thus,  necessarily the prior needs to be singular with respect to the $m$\text{-}dimensional Lebesgue measure. When the prior covariance of $\btheta$, $\bSigma_u$, is non-singular, such as in the standard Bayesian formulation of the FH model, it is not possible for the posterior of $\btheta$ to be supported on $\bTheta$, thereby making it unlikely that the posterior mean will belong to the set $\bTheta$.  

An advantage of building the constraints into the prior is that not only are point posterior estimates constrained, but the entire posterior is supported on the constrained set. Thereby, other exercises, such as uncertainty quantification via credible set construction, adhere to the constraints and no post-hoc adjustment is required.

We propose to assign a prior that is fully supported on $\bTheta$, leads to a posterior supported on $\bTheta$, and is proper with respect to the restricted Lebesgue measure on $\bTheta.$  
Specifically, let $\lambda_{\Theta}$  denote the restriction of $\lambda$, the Lebesgue measure on $\RR^m,$ to $\bTheta$ and define the measure, $\Pi_{\Theta},$ on $(\RR^m, \mcB(\RR^m)),$ that is fully supported on $\bTheta$ as follows: for every Borel set $A \in \mcB(\RR^m),$ let 
\be 
\Pi_{\Theta}(A) = \int_{A\bigcap \bTheta} \pi_{\Theta}(\btheta) d\lambda_{\Theta}(\btheta) 
\label{eq:prior_constrainted}
\ee
where the prior density $\pi_{\Theta}: \RR^m \to [0, \infty)$ is integrable with respect to $\lambda_{\Theta}$. If the measure is a proper probability measure, then in addition, we have
$ \Pi_{\Theta}(\RR^m) =    \int_{\bTheta}  \pi_{\Theta}(\btheta) d\lambda_{\Theta}(\btheta) = 1.$
We refer to the measure $\Pi_{\Theta}$ as a prior measure if in addition, the implied posterior is proper, i.e., $ \mcL(  \btheta|\bY)\pi_{\Theta}(\btheta)$ is integrable with respect to $\lambda_{\Theta}$, where $\mcL(\btheta | \bY)$ is the likelihood obtained from the model \eqref{eq:Basic_Fay_Herriot}. The posterior density $\pi_{\Theta}(\btheta | \bY)$ and the posterior measure are given by 
\be 
\Pi_{\Theta}(A | \bY) =  \int_{A \bigcap \bTheta}  \pi_{\Theta}(\btheta | \bY) d\lambda_{\Theta}(\btheta)  = \frac{\int_{A \bigcap \bTheta}  \mcL(\btheta | \bY)\pi_{\Theta}(\btheta) d\lambda_{\Theta}(\btheta) }{\int_{\bTheta}  \mcL(\btheta | \bY)\pi_{\Theta}(\btheta) d\lambda_{\Theta}(\btheta) }
\label{eq:poterior_constrainted}
\ee
for any $A \in \mcB(\RR^m)$ and $\int_{\bTheta}  d\Pi_{\Theta}(\btheta|\bY) = 1.$ If the constrained support $\bTheta$ is a convex set, as in the present application where $\bTheta$ is an affine subspace of $\RR^m,$ we can define the posterior mean in the usual manner as 
\be
E(\btheta | \bY) =  \int \btheta d\Pi_{\Theta}
(\btheta|\bY).
\label{eq:posterior_mean}
\ee
The following result then shows that the benchmarking constraints are enforced on the estimator $E(\btheta | \bY) $.
\begin{Proposition}
Let $\bTheta$ be as defined in \eqref{eq:parameter_space} and let $\Pi_{\Theta}(\btheta)$ and $\Pi_{\Theta}(\btheta|\bY)$  be the prior and posterior 
defined in \eqref{eq:prior_constrainted} and \eqref{eq:poterior_constrainted} corresponding to the likelihood $\mcL(\btheta | \bY)$  associated with  model \eqref{eq:Basic_Fay_Herriot}. Let $E(\btheta | \bY)$ be the posterior mean as defined in \eqref{eq:posterior_mean}.  Then 
\[    \bW^\top E(\btheta | \bY) = \hat{\btheta}_+.\]
\end{Proposition}
\begin{proof}
By the linearity of expectation and properties of the restricted measure
\[ \bW^\top E(\btheta | \bY) = \int \bW^\top \btheta d\Pi_{\Theta}
(\btheta|\bY)  = 
\int_{\bTheta}\hat{\btheta}_+ \pi_{\Theta}
(\btheta|\bY) d \lambda_{\Theta}(\btheta) = 
\hat{\btheta}_+ \Pi_{\Theta}(\RR^m|\bY)   = \hat{\btheta}_+.\]
\end{proof}
We will refer to any prior measure that satisfies the above formulation leading to a proper posterior probability measure supported on $\bTheta$ as a {\it benchmarking prior} and denote the posterior mean of $\btheta$ as ${\hat{\btheta}}^{BM},$ the benchmarked  estimator of $\btheta$.


\section{Construction of Gaussian Benchmarking  Priors}

In order to define a flexible class of restricted priors on $\bTheta$, we first seek to define a Gaussian measure on $\bTheta.$ Once a Gaussian measure is defined, other fully supported distributions on $\bTheta$ can be constructed as derivatives of the Gaussian measure, such as mixtures of normals on $\bTheta$ or coordinate-wise transformations.  
Moreover, we would like to use the Gaussian distribution for $\btheta$, conditional on the coefficients $\bbeta$ and the marginal variance $\sigma_u^2$,  as it facilitates conjugate prior specification in the FH model. Under the Gaussian prior, it is sufficient to identify the mean vector and the covariance matrix of the prior that will restrict the measure to the constrained set $\bTheta$. To build in similarity with the conventional FH prior specification, we demand the Gaussian measure has a mean that is linear in $\bX\bbeta$ and the prior variance of $\theta_i$ is a constant $\sigma_u^2$ for each $i.$ 
The following  result  characterizes the Gaussian  prior distribution. 

\begin{Proposition}
Consider a Gaussian random vector $\btheta \in \RR^m, $  $\btheta  \sim N_m(\bmu_{\theta}, \bSigma_u),$  with  mean  $\bmu_{\theta}$ and  variance $\bSigma_u$ depending on the parameters $\bbeta$ and $\sigma_u^2.$  Then $E(\btheta | \bbeta, \sigma_u^2) = \bX\bbeta + \bdelta$  where $\bdelta$ is such that $\bW^\top (\bX\bbeta + \bdelta) = \hat{\btheta}_{+},$    $var(\theta_i | \bbeta, \sigma_u^2) = \sigma_u^2 $  for $i = 1, \ldots, m, $ and $\btheta \in \bTheta$ with probability one if and only if  $\bSigma_{\bu} = \bsigma_u^2 \bR_u$ for some correlation matrix $\bR_u$ such that $\bW^\top \bR_u = \bzero,$ and $\bmu_{\theta} = \bX\bbeta + \bdelta_{\beta, \br}$ with 
\be 
\bdelta_{\bbeta,\br} =  \bW(\bW^\top\bW)^{-1}(\hat{\btheta}_{+} - \bW^\top\bX\bbeta) + \br
\label{eq:deltau}
\ee
for some vector $\br \in \ker(\bW^\top).$ 
\label{prop:benchmark_prior}
\end{Proposition}
\begin{proof}
It is straightforward to see that if $\bSigma_u = \bsigma^2\bR_u$ with $\bW^\top \bR_u = \bzero$ and if for some vector $\br \in ker(\bW^\top),$ $\mu_{\theta} = \bX\bbeta + \bdelta_{\bbeta, \br}$ with 
$\bdelta_{\bbeta,\br} =   \bW(\bW^\top\bW)^{-1}\bmu({\bbeta}) + \br$  then $\bmu({\bbeta}) = \hat{\btheta}_{+} - \bW^\top\bX\bbeta,$  and hence $\bW^\top \btheta = {\hat{\btheta}}_+$ with probability one. Moreover, from the choice of $\bSigma_u$ the variances of $\theta_i$ are all equal to $\sigma_u^2.$  Finally,  choosing $\bdelta = \bdelta_{\beta, \br}$ we have 
$E(\btheta | \bbeta, \sigma_u^2) = \bX\bbeta + \bdelta$ and   $\bW^\top (\bX\bbeta + \bdelta) = \hat{\btheta}_{+}.$

To see the converse, note that conditional on $\bbeta, \sigma_u^2,$ the condition $\btheta \in \bTheta$ w.p.1. implies that $Var(\bW^\top \btheta | \bbeta, \sigma_u^2) = \bzero  \leftrightarrow \bW^\top \bSigma_u\bW = \bzero \leftrightarrow \bSigma_u \bW = \bzero.$ Since the variances are constant and equal to $\sigma_u^2$  we have $\bSigma_u = \sigma_u^2 \bR_u$ for some correlation matrix $\bR_u$ satisfying 
\be
\bR_u \bW = \bzero.
\label{eq:Ru}
\ee
Moreover, the condition $\btheta \in \bTheta$ w.p.1.  implies that $E(\bW^\top \btheta) = \bW^\top \bmu_{\theta} = {\hat{\btheta}}_+. $
This implies
\[  \bmu_{\theta} =  \bW(\bW^\top \bW)^{-1} {\hat{\btheta}}_+ + (\bI - \bP_{\bW})\bu \]
for any $\bu \in \RR^m $   and $\bP_{\bW} = \bW(\bW^\top \bW)^{-1}\bW^\top.$ 
Let $\br = (\bI - \bP_{\bW})(\bu - \bX\bbeta).$  Since $\bu$ is  arbitrary, $\br$ can be any vector in  $\ker(\bW^\top).$ Writing $\bdelta_{\bbeta, \br}$ as in \eqref{eq:deltau}, we have the result. 
\end{proof}

This formulation is similar to the augmented model interpretation given in \cite{bell2013benchmarking}, but  the context is different and the  \cite{bell2013benchmarking} approach is primarily for providing  correct uncertainty quantification for internal benchmarking. In \cite{bell2013benchmarking}, the focus is to get the posterior mean to adhere to the benchmarking constraint whereas we want the entire prior and hence the entire posterior to be supported on the benchmarking hyperplane.

The Gaussian prior specification leads to the class of priors  
\be
\mcP_{\Theta} = \{\Pi_{\Theta}: \Pi_{\Theta}(\btheta | \bbeta, \sigma_u^2, \br, \bR_u)  \equiv  N_m(\bX\bbeta + \bdelta_{\bbeta, \br}, \bsigma_u^2 \bR_u) \}
\label{eq:prior_class}
\ee
where $\bdelta_{\bbeta,\br}$ is defined in \eqref{eq:deltau}. Note that each prior belonging to $\mcP_{\Theta}$ is a {\it Benchmarked prior} in the sense that they are fully supported on the benchmark constraint set $\bTheta$ and hence lead to posteriors that are fully supported on $\bTheta.$

\begin{Remark}
When the benchmarking constraints are observed without error—as in this article—the resulting posterior in the Gaussian case can be viewed as a limiting case of the formulations in \cite{hillmer1987benchmarking} and later \cite{durbin1997benchmarking}. In their framework, letting the measurement error variance of the constraint tend to zero forces the posterior to concentrate on the constraint set.
However, there is an important conceptual distinction. In \cite{hillmer1987benchmarking} and \cite{durbin1997benchmarking}, the constraint is incorporated through the likelihood by augmenting it with an additional term based on the linear constraint. As the error variance in the aggregate measurement shrinks to zero, this term converges to a Dirac mass over the constraint set, restricting the posterior to the constrained space.

In contrast, the proposed approach embeds the constraint directly in the prior, yielding a posterior that respects the constraint regardless of the likelihood specification. We find this formulation to be  direct and more flexible. 
In particular, it is possible to characterize different aspects of the prior specification process when constraints are built directly into the prior.
When constraints are imposed via the likelihood, prior propriety must still be ensured separately, whereas the prior-based construction handles this more naturally. Of course, the proposed approach only deals with  the case when the aggregates are observed without error. The approach of \cite{hillmer1987benchmarking}, on the other hand, allows for measurement error in the aggregate.  In the presence of measurement error the posterior is no longer supported on the constrained set, only the point estimate,  the posterior mean, obeys the constraints.   
\end{Remark}

To  specify a Gaussian prior of the type in \eqref{eq:prior_class},  conditional on $\bbeta$ and  $\sigma_u^2$,  we would need to specify hyperparameters $\br$ and $\bR_u$ that are defined through equations \eqref{eq:deltau} and \eqref{eq:Ru}. 
While the vector  $\br,$ which is any vector in $ker(\bW^\top),$ can be chosen relatively easily,  choosing the correlation matrix $\bR_u$ to satisfy restriction $\bW^\top\bR_u = \bzero$ is more involved. Fortunately, recent results by \cite{lafferty2025} provide a mechanistic approach to selecting correlations from the class of all correlation matrices that satisfy the constraint. The restriction on the correlation matrix amounts to selecting a correlation matrix $\bR_u$ such that all $q$ columns of $\bW$ are in the null space of $\bR_u.$ First, for a given $\bW$, we need to identify all $\bR_u$ that satisfy the constraints. The general problem of the existence of correlation matrices with specified vectors in the null space has been studied before by \cite{Barrett2003}. The problem is closely related to the ``nearest correlation matrix" problem studied by \cite{Higham1989,   Higham2002a, Higham2002b, Higham2016}. 
When $q = 1$, i.e., $\bW$ has a single column, say $\bw$, the problem has a surprisingly simple solution by \cite{Delorme1993}. For a  vector $\bw \in \RR^m$, there exists an $m\times m$ correlation matrix $\bR_u$ such that $\bR_u\bw = \bzero$, if and only if  $\bw$ is  {\it balanced} in the sense that it belongs to the set of balanced vectors 
\begin{equation}
    \mathscr{B}^m = \{ \bw \in \RR^m: 2\|\bw\|_{\infty} \leq \|\bw\|_1 \},
    \label{eq:balanced}
\end{equation}
where  for any vector $\bx \in \RR^m,$ $\|\bx\|_{\infty} = \sup\; |x_i|$ and $\|\bx\|_1 = \sum_{i=1}^m |x_i|.$ When $q > 1$, a general necessary and sufficient condition was not known.
Fortunately, if the constraint vectors have nonzero entries at different indices, as is the case in the benchmarking problem, checking the balance condition \eqref{eq:balanced} for each constraint becomes both necessary and sufficient. The solution for obtaining $\bR_u$ is based on the Projection onto Convex Sets (POCS) algorithm \cite{bauschke1993convergence, bauschke1997method}, which is a generalization of the popular alternating projection algorithm \cite{von1949rings, von1951functional, Rene2011} for finding solutions in the intersection of convex sets. Since the correlation matrices lie in the intersection of the following three convex sets,
\vspace{-.25in}
\beqa
    \mc{C}_1 &:=& \mathscr{S}^m_+ = \{ \bR_u \in \mathscr{S}^m :  \bR_u \succeq 0 \},  \nn \\
    \mc{C}_2 &:=& \mathscr{U}^m = \{ \bR_u \in \mathscr{S}^m : r_{ii} = 1, i = 1, \ldots,m\},\\
    \mc{C}_3 &:=& \mathscr{S}_{\bW} = \{ \bR_u \in \mathscr{S}^m : \bR_u\bW = \bzero \},
\eeqa
$\mathscr{S}^m$ being the set of symmetric $m \times m$ matrices, a solution to $\bR_u\bW = \bzero$ can be obtained by finding a member of the convex intersection $\mc{C} = \cap_{i=1}^3 \mc{C}_i.$ This is called a convex feasibility problem \cite{bauschke1996projection}, and a solution can be easily obtained using POCS by deriving the individual projections to $\mc{C}_1, \mc{C}_2$ and $\mc{C}_3.$
In the present application, the constraint vectors are the columns of the weight matrix $\bW$, and the entries are typically zeros and ones for simple aggregation.
Then each column satisfies the balanced condition \eqref{eq:balanced} and an $\bR_u$ with $\bR_u \bW = \bzero$ can be constructed in closed-form. For more general benchmarking constraints where the aggregations are weighted aggregations, the POCS algorithm has to be employed to obtain $\bR_u.$
\subsection{Gaussian Benchmarking Prior via Reparameterization}

The main challenge will be to specify a prior that is fully supported on $\bTheta$ and lends itself to reasonable posterior computations. Thus, given an $\bR_u$ such that $\bR_u\bW = \bzero$, it is easier to construct the prior and posterior through an equivalent parameterization on $\bTheta.$
Consider a reparametrization $\btheta\rightarrow\beeta(\btheta)$, such that 
\be 
\beeta := \beeta(\btheta) = \bM\btheta - \bdd, \quad \mbox{ where } \bM = \begin{bmatrix}
\bW^\top \\
\bK
\end{bmatrix},\bdd = \begin{bmatrix}
{\hat{\btheta}}_+\\
\bzero
\end{bmatrix}.
\label{eq:reparameterization}
\ee
We will choose $\bK$ such that $\bM$ is nonsingular and the mapping is a bijection, giving a proper reparameterization. In fact, we will choose the rows of $\bK$ to be orthonormal basis of $ker(\bW^\top)$, i.e., $\bK\bW = \bzero$ and $\bK\bK^\top = \bI_{m-q}.$ 
 Under the reparameterization $\btheta = \bM^{-1}(\beeta + \bdd).$
Under the constraint, the transformed parameters have the form $\beeta = (\bzero : \beeta_2)^\top$ where the first $q$ components of $\beeta$ are zero and the remaining $m - q$ components, $\beeta_2$ are free parameters. In other words, the transformed parameter space is 
\be
\bXi = \{ \beeta \in \RR^m: \beeta = [\bzero^\top: \beeta_2^\top]^\top \}.
\label{eq:transformed_parameter_space}
\ee 
\vspace{-1em}
Under the constraints, the area means are related to the transformed parameters as 
\be
 \btheta =  \bM^{(1)}\hat{\btheta}_+ +
 \bM^{(2)}\beeta_2
 \label{eq:eta_to_theta}
\ee
where $\bM^{-1} = [\bM^{(1)} : \bM^{(2)}]$ is a conformable partition. 
We will invoke a prior on $\bXi$ and thereby invoke a prior on $\bTheta$ via \eqref{eq:eta_to_theta}. If $\pi_{\eta}(\beeta)$
is a prior on $\bXi$ that yields a posterior such that 
 ${\hat{\beeta}}_2 = E(\beeta_2 | \bY)$ is the posterior mean of $\beeta_2$, then the benchmarked posterior estimate of $\btheta$ is 
\be
{\hat{\btheta}}^{BM} = \bM^{(1)}{\hat{\btheta}}_{+} + \bM^{(2)}{\hat{\beeta}}_2.
\label{eq:BM_estimate}
\ee
 Since $\bW^\top\bM^{(1)} = \bI$, and $\bW^\top\bM^{(2)} = 0$, it follows that $\bW^\top {\hat{\btheta}}^{BM} = {\hat{\btheta}}_{+}$.

To specify a prior on $\bXi$, we  adopt an approach which has high fidelity to the standard HB framework for the FH model, with the addition of the benchmarking constraints. Given $\bR_u$ such that $\bR_u\bW = \bzero$, consider the following modification of the standard HB prior  
\be
    \btheta|\bbeta, \sigma_u^2 \sim N(\bX\bbeta + \bdelta_{\bbeta, \br},\sigma_u^2 \bR_u), \;\; 
    \pi(\bbeta|\sigma_u^2) \propto 1, \;\; \sigma_u^2 \sim IG(a,b).
\label{eq:HB_constraint}    
\ee
where $\bdelta_{\beta, \br}$ is defined in \eqref{eq:deltau}.
It is straightforward to see the prior in \eqref{eq:HB_constraint} is fully supported on $\bTheta$ and satisfies the benchmarking constraints.
Under the new parametrization $\btheta\rightarrow\eta(\btheta)$, the HB model becomes
\beqn
    \mathbf{Y}|\beeta,\bbeta &\sim& N(\bM^{-1}(\beeta + \bdd),\bSigma_e), \nn \\
    \beeta|\bbeta, \sigma_u^2 &\sim& N({\bM \left(\bX \bbeta + \bdelta_{\bbeta, \br} \right) } -\bdd,\sigma_{u}^2\bM \bR_u\bM^\top), \nn \\
    \pi(\bbeta|\sigma_u^2) &\propto& 1, \quad \quad \sigma_u^2 \sim IG(a,b).
    \label{eq:HBNewparam}
\eeqn
Using the reparametrization, and \eqref{eq:HBNewparam},
\beqn
\bY | \beeta_2,\bbeta,\sigma_u^2 \sim N(\bM^{(1)}{\hat{\btheta}}_{+} + \bM^{(2)}\beeta_2,\bSigma_e), \qquad
\beeta_2|\bbeta,\sigma_u^2 \sim N(\bK\bX\bbeta + { \bK \br}, \sigma_u^2\bQ), 
\label{eq:HBeta2}
\eeqn where $\bQ = \bK \bR_u \bK^\top$. 
The joint posterior density of $(\beeta_2, \bbeta,\sigma_u^2)$ has straightforward full conditional distributions, facilitating sampling from the joint posterior via Gibbs sampler. The full conditionals are given by 
\beqa
\beeta_2 | \bY,\bbeta, \sigma_u^2 &\sim& N\left( {{\bmu}}_{\beeta_2}, \  {{\bSigma}}_{\beeta_2} \right), \nn \\
\bbeta \mid \bY,\beeta_2,\sigma_u^2 &\sim& N\left({{\bmu}}_{\beta},{{\bSigma}}_{\bbeta} \right), \nn \\
\sigma_u^2 |\bY,\bbeta, \beeta_2 &\sim& \text{IG}\left( \frac{m-{q}}{2} + a,    \frac{1}{2} \left( \beeta_2 - {\bK \left( \bX \bbeta + \br \right)} \right)^\top \bQ^{-1} \left( \beeta_2 - { \bK \left( \bX \bbeta + \br \right)} \right) + b \right),
\label{eq:postHB}
\eeqa
where 
\beqn
{{\bmu}}_{\eta_2} &=& {{\bSigma}}_{\eta_2} \left( (\bM^{(2)})^{T} \bSigma_e^{-1} ( \bY - \bM^{(1)} {\hat{\btheta}}_{+} ) + \frac{1}{\sigma_u^2}\bQ^{-1} \bK { \left( \bX \bbeta + \br \right)} \right), \nn \\
{{\bSigma}}_{\beeta_2} &=& \left( (\bM^{(2)})^{T} \bSigma_e^{-1} \bM^{(2)} + \frac{1}{\sigma_u^2}\bQ^{-1} \right)^{-1},\nn \\
{{\bmu}}_{\bbeta} &=& {\bSigma}_{\bbeta} \left(\bX^{T} \bK^{T}\bQ^{-1} { \left( \beeta_2 - \bK \br \right)} \right)/\sigma_u^2 , \nn \\
{{\bSigma}}_{\bbeta} &=& \sigma_u^2\left( \bX^{T} \bK^{T}{\bQ}^{-1}\bK\bX \right)^{-1}.
\lb{eq:gibbs}
\eeqn

\begin{Remark}
Under Assumption~1, we have $rank(\bK\bX) =  p  -  dim(\mcC(\bX) \bigcap \mcC(\bW)) = p.$ Hence the  matrix $\left( \bX^{T} \bK^{T}{\bQ}^{-1}\bK\bX \right)$ in  the expression for $\bSigma_{\bbeta}$ in  \eqref{eq:gibbs} is invertible.
\end{Remark}

Noting that $\bK \bM^{(1)} = 0$ and following \eqref{eq:BM_estimate}, we will have the benchmarked posterior estimate  
\[
{\hat{\btheta}}^{BM} = \bM^{(1)}{\hat{\btheta}}_{+} + \bM^{(2)}{\widehat{E(\beeta_2 | \bY)}}. 
\]
The reparameterization in \eqref{eq:reparameterization} works because $\beeta_1$ becomes degenerate at $\bzero$ under the choice of $\bR_{\bu}$ satisfying \eqref{eq:Ru}. However,  $\beeta_2$ remains the free variable. The following result shows that the posterior of $\btheta$ obtained using the reparameterization is invariant to the choice of $\bK.$
\begin{Proposition}
Consider the reparameterized variable of interest $\btheta = \bM^{(1)}{\hat{\btheta}}_{+} + \bM^{(2)}\beeta_2 $ obtained via the reparameterization \eqref{eq:reparameterization}. The posterior of $\btheta$ is invariant  to the choice of $\bK$ in \eqref{eq:reparameterization}.
\end{Proposition}
\begin{proof}
To see this,  consider the group action $\bK \to \bB\bK$ for some nonsingular $\bB$. Then it is easy to see that under this group action,  the reparameterization gives a new $\bM$ matrix such that $\bM^{(2)} \to  \bM^{(2)}\bB^{-1}$ and $\beeta_2 \to  \bB\beeta_2.$  Hence the posterior of $\btheta =  \bM^{(1)}{\hat{\btheta}}_{+}  + \bM^{(2)}\beeta_2$ remains invariant.
\end{proof}

\subsection{Max-Fidelity Benchmarking Prior
}
The formulation of the fully supported Gaussian prior provides a Bayesian posterior for $\btheta$ that is fully supported on $\bTheta$ and hence satisfies the benchmarking constraints. For any choice of $\br \in ker(\bW^\top)$ and $\bR_u$, as given in \eqref{eq:deltau} and \eqref{eq:Ru},  the prior $\Pi_{\Theta}(\btheta| \bbeta, \bsigma_u^2 \br, \bR_u) = N_m(\bX\bbeta + \bdelta_{\bbeta, \br}, \sigma_u^2\bR_u)$ leads to a fully supported posterior. While each such choice leads to a different estimator of $\btheta$ that satisfies the benchmarking constraint,  here we provide a principled way of choosing $\br$ and $\bR_u.$ 
In the original  FH formulation the prior for $\btheta$ only depends on $\bbeta$ and $\sigma_u^2$ and noninformative  priors are placed on $\bbeta$ and $\bsigma_u^2.$ In the present situation, the benchmarking constraints    forces introduction of $\br$ and $\bR_u.$  We choose the hyperparameters to have maximum fidelity to the original FH prior formulation in the sense that  we choose $\br$ and $\bR_u$ to minimize the Kullback-Leibler (KL) divergence between the benchmarking prior $N(\bX\bbeta + \bdelta_{\beta, \br}, \sigma_u^2 \bR_u)$ and the prior in the original FH formulation $N(\bX\bbeta, \sigma_u^2 \bI).$ In view of (\ref{eq:eta_to_theta}), it is enough to specify a max-fidelity prior on $\beeta_2$. Under the unconstrained FH model, the implied prior on  $\beeta_2$ would be $\mathcal{N}\!\left(\bK \bX \bbeta,\,\sigma_{u}^{2}\,\bI_{m-q}\right).$
Let 
\be
D_{\mathrm{KL}} = D_{\mathrm{KL}}\!\left(
    \mathcal{N}\!\left(\bK \bX \bbeta + \bK \bdelta_{\bbeta, \br},\,
                       \sigma_{u}^{2}\,\bK \bR_u \bK^{T}\right)
    \;\Big\|\;
    \mathcal{N}\!\left(\bK \bX \bbeta,\,\sigma_{u}^{2}\,\bI_{m-q}\right)
  \right)
\lb{eq:KL}
\ee
be the KL divergence between the two sets of priors for $\beeta_2$, the benchmarking prior  given  $\bbeta, \sigma_u^2, \br, \bR_u$ and the original FH prior  given $\bbeta, \sigma_u^2.$
Then the following proposition gives the values of $\br$ and $\bR_u$ for the benchmarking prior to have maximum fidelity to the original FH prior.

\begin{Proposition}
Consider the FH model as defined in \eqref{eq:HB} and let the benchmarking prior with respect to the benchmarking constraint set $\bTheta$ be as given in Proposition~\ref{prop:benchmark_prior}. 
Let $\bK =  [\bk_{*1} : \cdots : \bk_{*,m}]$ be any $(m-q) \times m$ matrix  whose rows form an  orthonormal  basis of $\ker(\bW^\top),$ i.e. $\bK\bW = \bzero$ and $\bK\bK^\top = \bI_{m-q}.$ Then $\br$ and $\bR_u$  minimize the KL divergence  
$D_{\mathrm{KL}} $, as defined in \eqref{eq:KL}, if and only if  $\br = \bzero$ and $\bR_u$ is equal to $\bR_u^* = \bK^\top \bB^* \bK$ where $\bB^*$ is a solution to 
\begin{equation}
\bB^{\star} = \underset{\bB \in \mcC_{\bB}} \argmin \;  [\tr(\bB) - \log\det (\bB)]
  \label{eq:B_problem}
\end{equation}
with $\mcC_{\bB} = \{ \bB \in \mathbb R^{(m-q)\times(m-q)}, \bB \succ \bzero,\; \bk_{*i}^\top\bB\bk_{*i} = 1,\;  i = 1, \ldots, m\}.$
Moreover, the solution $\bR_u^\star$ does not depend on the particular choice of $\bK.$
If the feasible set is nonempty, the solution $\bR_u^\star$ is unique. 
\label{prop:max_fidelity}
\end{Proposition}

\begin{proof}
It is straightforward to see that the expression for $D_{\mathrm{KL}}$ in \eqref{eq:KL} reduces to
\[ D_{\mathrm{KL}}  = \frac{1}{2}[\tr(\bK\bR_u\bK^\top) -\log\det(\bK\bR_u\bK^\top)  + \sigma_u^{-2}|| \bK\br||_2^2 - m + q]. \]
In terms of $\br$ the expression is minimized iff $\|\bK\br\| = 0$. Since $\br \in \ker(\bW^\top)$ we have $\|\bK\br\| = \|\br\|.$ Hence the optimal choice for $\br$ is $\br = \bzero.$  Thus, the optimization problem for finding the optimal $\bR_u$ reduces to 
\begin{equation}
\bR_u^{\star} = \underset{\bR_u \in \mathscr{C}^m, \bR_u\bW = \bzero} \argmin \;  [\tr(\bK\bR_u\bK^\top) - \log\det (\bK\bR_u\bK^\top)],
  \label{eq:reduced}
\end{equation} 
where $\mathscr{C}^m$ is the space of all $m\times m$ correlation matrices. Since $\bR_u\bW = \bzero$ and $\bR_{u}$ is symmetric, every column of $\bR_u$ lies in $\ker(\bW^\top)$. As the rows of $\bK$ form an orthonormal basis of $\ker(\bW^\top)$, we have $\bK^\top\bK = \bI_m - \bP_W$, the orthogonal projector onto $\ker(\bW^\top).$ Hence $(\bK^\top\bK)\bR_u(\bK^\top\bK) = \bR_u. $ With $\bK\bK^\top = \bI_{(m-q)\times (m-q)}$, this is equivalent to $\bR_u  = \bK^\top\bB\bK $ where $\bB = \bK\bR_u\bK^\top \in\mathbb{R}^{(m-q)\times(m-q)}$ is positive semidefinite. 

With this notation, the optimization in ~\eqref{eq:reduced} reduces to solving 
\begin{equation}
\bB^{\star} = \underset{\bB \in \mcC_{\bB}} \argmin \;  [\tr(\bB) - \log\det (\bB)].
  \label{eq:B_problem}
\end{equation} 
Problem~\eqref{eq:B_problem} is strictly convex. Thus, the KKT conditions are
necessary and sufficient. Let $\bmu = (\mu_1, \ldots, \mu_m)^\top$ be Lagrangian multipliers. The Lagrangian can be written as 
\beqa  
L(\bB, \bmu) &=&  \tr(\bB) - \log\det (\bB) + \sum_{i=1}^m  \mu_i(\bk^{\top}_{*i}\bB\bk_{*i} - 1) \\
&=&  \tr(\bB(\bI + \sum_{i=1}^m \mu_i \bk_{*i}\bk^\top_{*i})) - \log\det (\bB) - \sum_{i=1}^m \mu_i.
\eeqa
Differentiating with respect to $\bB$ 
we get the stationarity condition as 
\[ \bI_{m-q} + \sum_{i=1}^m \mu_i \bk_{*i}\bk^\top_{*i} - \bB^{-1} = 0 \]
which implies
$\bB^{-1} = \bI_{m-q} + \bK\,\diag(\bmu)\,\bK^\top$.
The diagonal
constraints $\bk^\top_{*i}\bB\bk_{*i} = 1, i = 1, \ldots, m, $ determine the multipliers $\bmu^{\star}$ from the set $m$ of nonlinear equations given by 
\beqn \bk^{\top}_{*i}(\bI_{m-q} + \bK \diag(\bmu)\bK^\top)^{-1}\bk_{*i} =1, \; i = 1, \ldots, m. \label{prop4_nonlineqn} \eeqn
If  $\bmu$ and $\blambda$  are two sets of multipliers, then at solution we have $\bK(\diag(\bmu) - \diag(\blambda))\bK^\top = 0$ and hence $\sum_{i=1}^m (\mu_i - \lambda_i)\bk_{*i}\bk^\top_{*i} = 0.$ This implies $\mu_i = \lambda_i$, i.e. the multipliers are uniquely determined only if $\bk_{*i}\bk^\top_{*i}$ are linearly independent of each other. 
\end{proof}


The nonlinear equations in \eqref{prop4_nonlineqn} can be solved efficiently 
using Newton's method. The Jacobian admits the closed-form expression with $(i,j)$th element taking the form
$J_{ij} = - (R_{ij})^2$, enabling fast second-order updates.

Based on Proposition~\ref{prop:max_fidelity}, the vector $\br$ can be chosen as $\br = \bzero$ for the max-fidelity choice. The choices for $\bR_u$ given as $\bR_u^*$, defined in the statement of Proposition~\ref{prop:max_fidelity}. Thus, the {\it max-fidelity  benchmarking prior} is given as $\btheta|\bbeta, \sigma_u^2 \sim N((\bI - \bP_{\bW})\bX\bbeta + \bW(\bW^\top\bW)^{-1}{\hat{\btheta}}_+, \sigma_u^2 \bR_u^*)$  with $\bR_u^*$ satisfying \eqref{eq:reduced}. 

\begin{Corollary}
\label{cor: w1opt}
If $\bW = \bone_m$, then
\[
\bR_u^\star
=
\frac{m}{m-1}\left(\bI_m - \frac{1}{m}\bone_m \bone_m^\top\right),
\]
i.e., $\bR_u^\star$ is the intraclass correlation matrix with off-diagonal entries $-1/(m-1)$.
\end{Corollary}
\begin{proof}
If $\bW =\bone_m$, then
$
\ker(\bW^\top)=\ker(\bone_m^\top)
=\{\bx\in\RR^m:\bone_m^\top \bx=0\}.
$
Let $\bK^\top \in\RR^{m\times(m-1)}$ have orthonormal columns spanning this space. Then
\[
\bK^\top\bK=\bI_m-\frac{1}{m}\bone_m\bone_m^\top.
\]
By Proposition~\ref{prop:max_fidelity}, the optimizer satisfies
\(
\bR_u^\star=\bK^\top\bB^\star\bK
\)
for some $\bB^\star\succ \bzero$. Since the problem is invariant under permutations of the coordinates, uniqueness implies that $\bR_u^\star$ must be exchangeable, and hence
\[
\bB^\star=c\,\bI_{m-1}
\]
for some constant $c>0$. Therefore
\[
\bR_u^\star
=
c\,\bK^\top\bK
=
c\left(\bI_m-\frac{1}{m}\bone_m\bone_m^\top\right).
\]
Using the constraint $\diag(\bR_u^\star)=\bone_m$, gives $c = \frac{m}{m-1},$ yielding the result.
\end{proof}





\section{Benchmarking for Robust Fay-Herriot Model}
There are  extensions of the basic FH model that accommodate the possibility of outliers in the area level means. 
The residuals for the $i$-th area, in the standard FH model, can be expressed as the sum of sampling error $e_i$ and area-level random effect $u_i$. 
A large value of $\vert u_i\vert$ or $\vert e_i\vert$ could cause an outlier in the residuals. However, in practice, it is difficult to determine whether such an outlier originates from the sampling error component or the area-level random effect (\cite{bell2006trob}). To mitigate the influence of outliers, several robust extensions of the Fay–Herriot model have been proposed in the literature (see, e.g., \cite{datta1995robust,bell2006trob,fabrizi2010robust}). \cite{datta1995robust} introduced robustness by modeling the random effects through a scale–normal mixture, ensuring propriety of the posterior under broad conditions. \cite{bell2006trob} proposed a related formulation using a multivariate $t$ prior with fixed degrees of freedom to accommodate extreme observations. \cite{fabrizi2010robust} further proposed a framework by employing an exponential power prior for the area-level random effects. \par
For this work, we proceed under the assumption that the outliers are introduced due to large values of the area-level random effect $\vert u_i\vert$. We adopt and extend the framework introduced by \cite{bell2006trob} to enforce benchmarking at an aggregated level. The Gaussian random–effects prior is replaced with a multivariate-$t$ prior
\be
    \btheta \mid \bbeta,\sigma_u^2 \sim t_\nu\!\left(\bX\bbeta + \bdelta_{\beta},\ \frac{\nu-2}{\nu}\sigma_u^2 \bR_u\right), \;\;
    \pi(\bbeta\mid\sigma_u^2) \propto 1, \;\; \sigma_u^2 \sim IG(a,b),
\label{eq:tHB_constraint}
\ee
where $\bdelta_{\beta}, \bR_u$ are subject to similar benchmarking restrictions as in \eqref{eq:deltau} and \eqref{eq:Ru}, where $\nu >2$.
The above HB model can be conveniently represented by expressing the multivariate $t$ distribution as a normal-gamma scale mixture, as follows.
\beqn
\btheta\mid \bbeta,\sigma_u^2,\lambda &\sim& N\!\left(\bX\bbeta + \bdelta_{\beta},\ \frac{\sigma_u^2}{\lambda}\,\bR_u\right), \qquad
\lambda \sim \text{Gamma}\!\left(\frac{\nu}{2},\frac{\nu-2}{2}\right),
\label{eq:t_mixture}
\eeqn
which is fully supported on $\bTheta$. Here, $\text{Gamma}(\alpha,\beta)$ denotes a Gamma distribution with shape $\alpha$ and rate $\beta$. Under above formulation $Var(\btheta|\bbeta,\sigma_u^2) = \sigma_u^2 \bR_u .$
Then, under the reparametrization $\btheta\mapsto\beeta(\btheta)$, the HB model becomes
\vspace{-.1in}
\beqn
    \bY\mid\beeta,\bbeta &\sim& N\big(\bM^{-1}(\beeta+\bdd),\,\bSigma_e\big), \nn\\
    \beeta\mid\bbeta,\sigma_u^2,\lambda &\sim& N\!\big(\bM(\bX\bbeta + \bdelta_{\beta})-\bdd,\ \tfrac{\sigma_u^2}{\lambda}\,\bM\bR_u\bM^\top\big), \nn\\
    \pi(\bbeta\mid\sigma_u^2) &\propto& 1, \;\; 
    \sigma_u^2 \sim IG(a,b), \;\; \lambda \sim \text{Gamma}\!\left(\tfrac{\nu}{2},\tfrac{\nu-2}{2}\right).
\label{eq:tHBNewparam}
\eeqn Projecting onto the benchmark-free coordinates,
\vspace{-1em}
\beqn
\bY \mid \beeta_2,\bbeta,\sigma_u^2,\lambda &\sim& N\!\left(\bM^{(1)}\hat{\btheta}_+ + \bM^{(2)}\beeta_2,\ \bSigma_e\right), \nn\\
\beeta_2 \mid \bbeta,\sigma_u^2,\lambda &\sim& N\!\left(\bK\bX\bbeta,\frac{\sigma_u^2}{\lambda}\,\bQ\right), \qquad
\bQ = \bK\bR_u\bK^\top .
\label{eq:tHBeta2}
\eeqn
The above formulation lends itself to a closed-form full conditional distributions for all parameters. Hence, the joint posterior of $(\beeta_2,\bbeta,\sigma_u^2,\lambda)$ can be sampled using a  Gibbs sampler. The full conditional distributions are provided in the supplementary material.

\section{Simulation Study}
\label{sec:Simulation}

We conducted a Monte Carlo simulation study with $B=500$ replicates to empirically evaluate the performance of our proposed approach. Because the proposed approach guarantees that all posterior draws satisfy the benchmarking constraints, it also enables the construction of $95\%$ credible intervals for $\theta$ with the constrained set.
To assess the performance of our proposed benchmarked estimator, we study the mean squared error (MSE), the coverage probability (CP), the average length (AL) and the interval score (IS) \citep{Gneiting01032007} of the credible intervals. For a nominal $(1-\alpha)$ prediction interval $[L,U]$ and a realized value $\theta$, the interval score is defined as
\(
IS_\alpha(L,U;\theta) = (U-L) + \frac{2}{\alpha}(L-\theta)\Ind(\theta < L) + \frac{2}{\alpha}(\theta-U)\Ind(\theta > U).
\label{eq:interval_score}
\), where $\Ind(\cdot)$ denotes the indicator function. This score rewards narrow intervals through the width term $(U-L)$ and penalizes miscoverage at both ends of the confidence interval, with the penalty scaled by $\frac{2}{\alpha}$ and proportional to the amount of miscoverage. 

For comparison, we considered the Bayesian benchmark estimator developed in Theorem 1 of \cite{datta2011bayesian} (hereafter DGSM), given by $\hat{\theta_i}^{dgsm} = \hat{\theta}_i + \frac{w_i}{s\phi_i}(\hat{\theta}_+-\sum_{i=1}^{m}w_i\hat{\theta}_i)$, where $s=\sum_{i=1}^{m}w_i^2/\phi_i$, with $\phi_i = D_i^{-1}$ and $w_i = 1$. However, as a point estimator, the DGSM estimator does not provide uncertainty quantification. Additionally, we include the unconstrained FH models for each setup for comparison, which we will subsequently refer to as UHB.

We consider three simulation setups to assess the performance of our proposed estimator under both correct and incorrect model specification. Specifically, we consider a correctly specified Gaussian Fay-Herriot model, a misspecified Gaussian FH model in which the fitted model omits relevant covariates, and a second misspecified model in which the fitted model both omits relevant covariates and includes extraneous ones. In each setup, we vary the number of small areas $m \in \{20,40,100\}$ to examine how the relative performance of the estimators changes with $m$.


Across each simulation setting, we generate $ p=7$ auxiliary covariates for each of the $m$ areas. The columns of the true covariate matrix $\bX$ are generated independently as $X_1\sim N(0,1)$, $X_2\sim U(0,10)$, $X_3\sim\mathrm{Gamma}(2,1)$, $X_4\sim\mathrm{Bernoulli}(0.5)$, $X_5\sim U(0,5)$, $X_6\sim\mathrm{Gamma}(5,1)$, and $X_7\sim N(10,2^2)$ and kept fixed across all setups.  The true area-level means are generated as
$$
\btheta = \bX\bbeta + \bu, \qquad \bu \sim \mathcal{N}(0,\, 3^2\, \bI_m),
$$ with true regression coefficients $\bbeta = (0.05,\ 0.1,\ 0.5,\ -0.5,\ 0.01,\ 0.6,\ 0.3)^\top$ used across all setups. At each replicate, a new realization of the direct estimates $\bY$ is generated as $\bY \sim \mathcal{N}(\btheta, \bD)$, with  the sampling variances of the direct estimates, $\bD = \mathrm{diag}(D_1, \ldots, D_m)$,   generated once as  iid values from $\mathrm{Uniform}[1,30]$, and held fixed across replicates. 
\begin{itemize}
    \item \textbf{Setup I (correctly specified):} The fitted model is given the full covariate matrix $\bX \in \mathbb{R}^{m\times p}$ used to generate $\btheta$. 

    \item \textbf{Setup II (omitted-variable misspecification):} The fitted model is given only the first four columns of $\bX$, omitting $x_5, x_6$ and $x_7$ (with true coefficients $\beta_5 = 0.01,\beta_6 = 0.6$, $\beta_7 = 0.3$) from the design matrix, so that the fitted model underfits relative to the true DGP.
    
    \item \textbf{Setup III (omitted-and-added-variable misspecification):} the fitted model again omits $x_5,$ $x_6$ and $x_7$, however it includes two spurious covariates $z_1 {\sim} \mathcal{N}(2,1)$ and $z_2 {\sim} \mathcal{N}(4,1)$ that play no role in $\btheta$.
\end{itemize}


To impose the benchmarking constraints, we use a single benchmarking vector $\bW = \boldsymbol{1}_m$ to simulate a benchmark total for the area-level population parameters at the aggregate level. 
Following the formulation of the article, we assume that  the value at the aggregate level is observed with little to negligible sampling error based on external information, either a larger survey or a census. Hence, we consider benchmarking the area-level estimators against the total of the true area-level mean vector $\btheta$. 
Since $\bW = \mathbf{1}_m$, to extend $\bW^\top$ to a nonsingular matrix, $\bM$, we choose $\bM$ to be the Discrete Cosine transform matrix (DCT) \citep{strang1999discrete} with $\bW^\top$ as its first row. The extension matrix $\bK$ thus comprises rows 2 through $m$ of the DCT matrix.


To enforce benchmarking constraints, we construct benchmarking priors following the methodology proposed for the Gaussian FH model. For implementation of the proposed method, we first find the correlation matrix  $\bR_u$ such that $\bR_u\mathbf{1} = 0$ and keep it fixed throughout the simulation. It turns out from Corollary~\ref{cor: w1opt}, that the max-fidelity choice for $\bR_u$ would be the {\it intraclass correlation matrix} with the intraclass correlation value as $\frac{-1}{m-1}$. All models were implemented in R \citep{r2024r}. Posterior inference used 30,000 posterior draws after discarding 20,000 initial samples as burn-in. 

\begin{table}[ht]
\centering
\small
\setlength{\tabcolsep}{3pt}
\caption{MSE across Setup I (correctly specified), Setup II (omitted-variable misspecification), and Setup III (omitted-and-added-variable misspecification), for the Direct estimator, unconstrained hierarchical Bayes (UHB), DGSM, and the proposed constrained benchmarking prior based estimator (CBP). The number of replications is  500.}
\begin{tabular}{l rrr ||rrr ||rrr}
\toprule
 & \multicolumn{3}{c}{\textbf{Setup I}} & \multicolumn{3}{c}{\textbf{Setup II}} & \multicolumn{3}{c}{\textbf{Setup III}} \\
\cmidrule(lr){2-4} \cmidrule(lr){5-7} \cmidrule(lr){8-10}
\textbf{Model} & $m{=}20$ & $m{=}40$ & $m{=}100$ & $m{=}20$ & $m{=}40$ & $m{=}100$ & $m{=}20$ & $m{=}40$ & $m{=}100$ \\
\midrule
Direct & 16.69 & 15.90  & 16.04 & 16.48  &  16.14 & 16.24 & 16.28 & 16.12  & 16.13  \\
UHB   & 9.94 & 7.60 & 6.05 & 10.84 & 8.88  & 7.98  & 10.62 & 8.08  &  7.02  \\
DGSM   &  9.19& 7.25 & 5.99 & 10.01 & 8.34  & 7.64  & 9.95  & 7.74  & 6.84  \\
CBP    & 9.56 & 7.47 & 5.99 & 8.30  & 7.23  & 6.61 & 9.36  & 7.67  & 6.74  \\
\bottomrule
\end{tabular}
\label{tab:sim_mse_500}
\end{table}

\begin{figure}[ht]
    \centering
    \includegraphics[width=.85\linewidth]{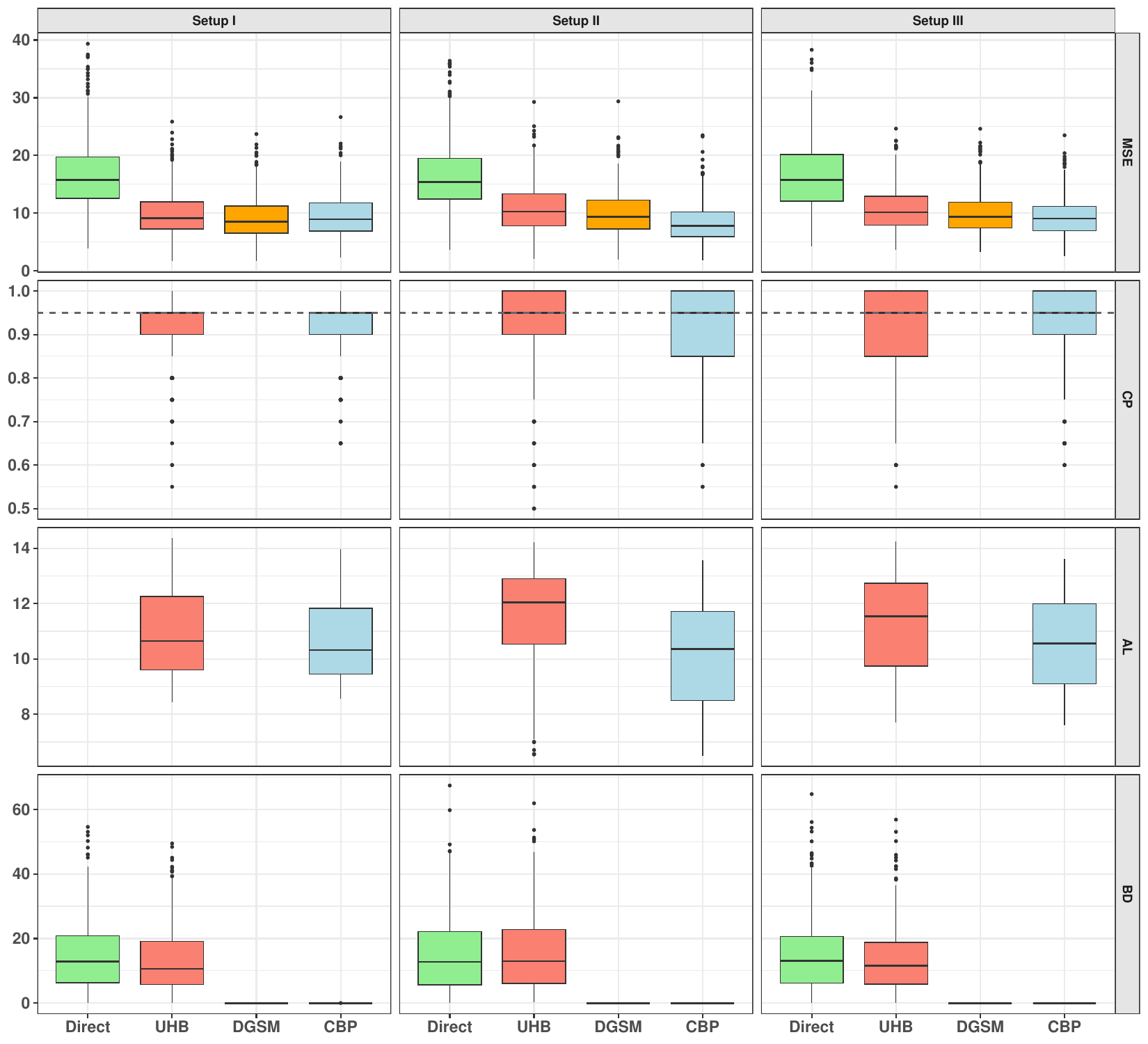}
    \caption{Boxplots of MSE, coverage probability (CP), average length (AL) of credible interval, and aggregate benchmarking discrepancy (BD) 
    for the direct estimator (green), unconstrained hierarchical Bayes estimator (red), DGSM estimator (orange), and proposed constrained benchmarking prior (blue) across the setups I, II \& III, based on $B = 500$ replicates.}
    \label{fig:Sim_Result_500_plot}
\end{figure}
For the standard FH model based estimates 
Figure~\ref{fig:Sim_Result_500_plot} presents a panel of comparative boxplots of four performance metrics (MSE, frequentist empirical coverage, average Length of the posterior credible intervals, and benchmarking discrepancy). The number of small areas, $m=20$, is the same across all three setups. Each row corresponds to a metric, and each column to a setup. The first row corresponds to the full simulation distribution of the MSE, which seconds the pattern we observed in Table~\ref{tab:sim_mse_500}. The second and third rows show comparative boxplots of frequentist coverage and the average length of the posterior credible intervals for UHB and CBP. As shown, CBP provides nearly comparable coverage with shorter credible intervals. Finally, the last row of the panel shows the benchmarking discrepancy: how much the estimates differ when aggregated from the true aggregate-level estimate. The figure shows that both exact benchmarking methods, DGSM and CBP, satisfy the exact benchmarking constraints, while the Direct and UHB estimates, when aggregated, deviate significantly from the aggregate-level Direct estimate. 

Table~\ref{tab:sim_mse_500} shows a consistent pattern across all nine setups. As expected, the direct estimator has  significantly larger MSE than the model-based estimators. Among the model-based estimators, the proposed CBP estimator consistently outperforms the UHB  and the DGSM estimators in terms of MSE; however, the margin varies across setups. Under correct specification (Setup I), CBP outperforms the UHB estimator by 0.99-3.8\% across different choices of $m$, with the margin being highest at $m=20.$ However, it is marginally outperformed by DGSM (0.7–4\%) in this particular instance.  This is not surprising since under correct model specification, there is no systematic bias for the benchmarking to correct, and hence the gain  in terms of bias reduction cannot compensate the increased variance. Under omitted-variable misspecification (Setup II), CBP is particularly advantageous compared to its counterparts. It reduces MSE by 17–24\% relative to UHB and 13–17\% relative to DGSM across all $m$.The systematic bias introduced from the omission of $x_5,$ $x_6$ and $x_7$ gets substantially corrected  via the exact benchmarking, providing CBP with a substantial gain in terms of MSE. The bias correction  propagates  more effectively for CBP through constrained prior specification compared to that for DGSM. Setup III (misspecified with omitted and added covariates) shows the same pattern as Setup II, but it is weaker. The CBP reduces MSE compared to UHB by 3.9\% to 11.9\% across $m=20,40,100$, whereas the gain with respect to DGSM is more modest with the highest gain (about 6\%) being  observed for $m = 20$. 
\begin{table}[ht]
\centering
\small
\setlength{\tabcolsep}{3pt}
\caption{Interval Score across Setup I (correctly specified), Setup II (omitted-variable misspecification), and Setup III (omitted-and-added-variable misspecification), for unconstrained hierarchical Bayes (UHB) and the proposed constrained benchmarking prior (CBP), based on 500 replications. $\alpha = 0.05.$}
\begin{tabular}{l rrr|| rrr ||rrr}
\toprule
 & \multicolumn{3}{c}{\textbf{Setup I}} & \multicolumn{3}{c}{\textbf{Setup II}} & \multicolumn{3}{c}{\textbf{Setup III}} \\
\cmidrule(lr){2-4} \cmidrule(lr){5-7} \cmidrule(lr){8-10}
\textbf{Model} & $m{=}20$ & $m{=}40$ & $m{=}100$ & $m{=}20$ & $m{=}40$ & $m{=}100$ & $m{=}20$ & $m{=}40$ & $m{=}100$ \\
\midrule
Direct & 18.482 &  18.022 &  18.029 & 18.295 & 17.918 & 18.153 & 18.523 & 17.938 & 17.986 \\
UHB &15.184 &  14.037 &  11.660 & 16.263 & 14.003 & 13.108 & 16.422 & 13.966 & 12.382 \\
CBP & 14.800  & 13.751 &  11.606 & 14.297 & 13.813 & 12.138 & 15.227 & 13.924 & 12.306 \\
\bottomrule
\end{tabular}
\label{tab:sim_is_500}
\end{table}

Table \ref{tab:sim_is_500} reports the interval scores for the unconstrained hierarchical Bayes (UHB) estimator and the proposed constrained benchmarking prior (CBP) across the three simulation setups and three sample sizes $(m = 20, 40, 100).$ Two clear patterns can be discovered from Table~\ref {tab:sim_is_500}. First, within each setup, the interval score decreases as the number of small areas increases. Second, and more central to our contribution, is that CBP achieves lower IS compared to UHB across all setups. In Setup~I, the improvement is modest since the model is correctly specified. However, the gains in terms of the IS are  significant in Setup~II and Setup~III.

\section{Real Data Analysis}
Small area estimation of poverty rates among school-going children is an important application of the SAE methodology, as these rates serve as a vital component in several policy decisions regarding fund allocation at both local and federal levels. We analyze data from the Small Area Income and Poverty Estimates (SAIPE) program of the U.S. Census Bureau. SAIPE provides model-based estimates of the proportion of children aged 5--17 living in poverty for the nation as a
whole, as well as different geographical levels such as states, counties, and school districts. Since these rates play a pivotal role in policymaking, the rates are often required  to be benchmarked against the aggregate level estimates. For this application, the state estimates were benchmarked to the direct estimate of the national  child poverty rate obtained from the  Annual Social and Economic Supplement of the Current Population Survey (CPS ASEC). We include four covariates alongside an intercept term: the food stamp participation rate, the IRS non-filer rate, and the IRS-based child poverty rate (each expressed as a percentage on a 0--100 scale), along with a residual term derived from fitting an analogous model to the 1990 Census long-form poverty rate estimates. The set up is similar to the analysis given in \cite{datta2011bayesian}.

The SAIPE program adopts an area-level unconstrained FH model for producing state-level model-based estimates \citep{datta2011bayesian}. We model the state-level direct estimates corresponding to the year 1999 using the proposed methodology (CBP) to accommodate the benchmarking constraints, $\bW^{\top}\hat{\btheta}_{BM} = \hat{\btheta}_{+},$ where $\bW$ is a column vector. Here $q =1,$ and each entry of $\bW$, $w_i$, is proportional to the population estimates of the number of school-going children that are eligible to be assigned a poverty status. The corresponding correlation matrix $\bR_{u}$ is obtained by solving the max-fidelity optimization problem in Proposition~\ref{prop:max_fidelity}. As in the simulation study, posterior inference was carried out via
Gibbs sampling from the full conditional distributions. We ran
50{,}000 iterations, discarding the first 20{,}000 as burn-in, which
yields 30{,}000 retained posterior draws.
\begin{figure}[H]
    \centering
    \begin{subfigure}[t]{0.48\linewidth}
        \centering
        \includegraphics[width=\linewidth]{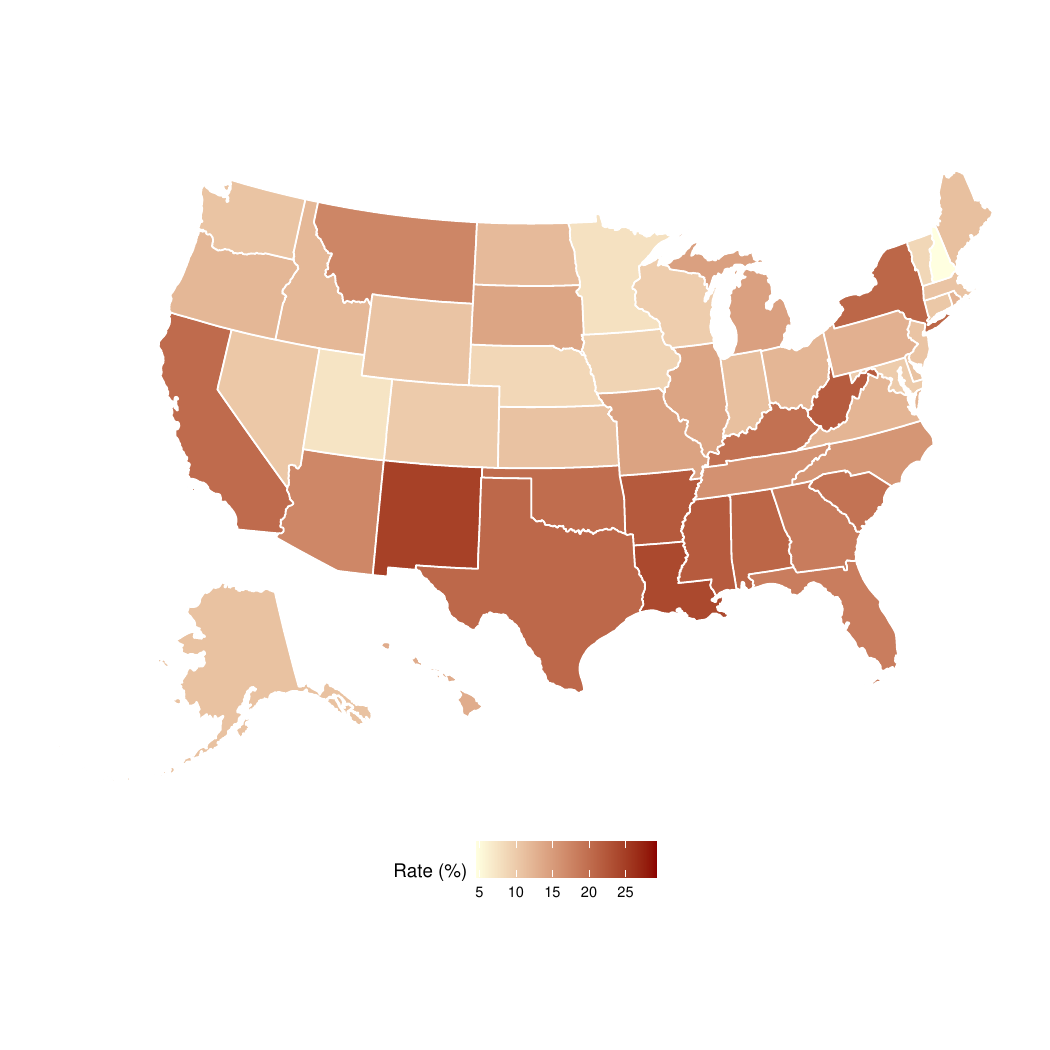}
        \caption {Poverty Rate Estimates}
        \label{fig:estimate_map_plot}
    \end{subfigure}
    \begin{subfigure}[t]{0.48\linewidth}
        \centering
        \includegraphics[width=\linewidth]{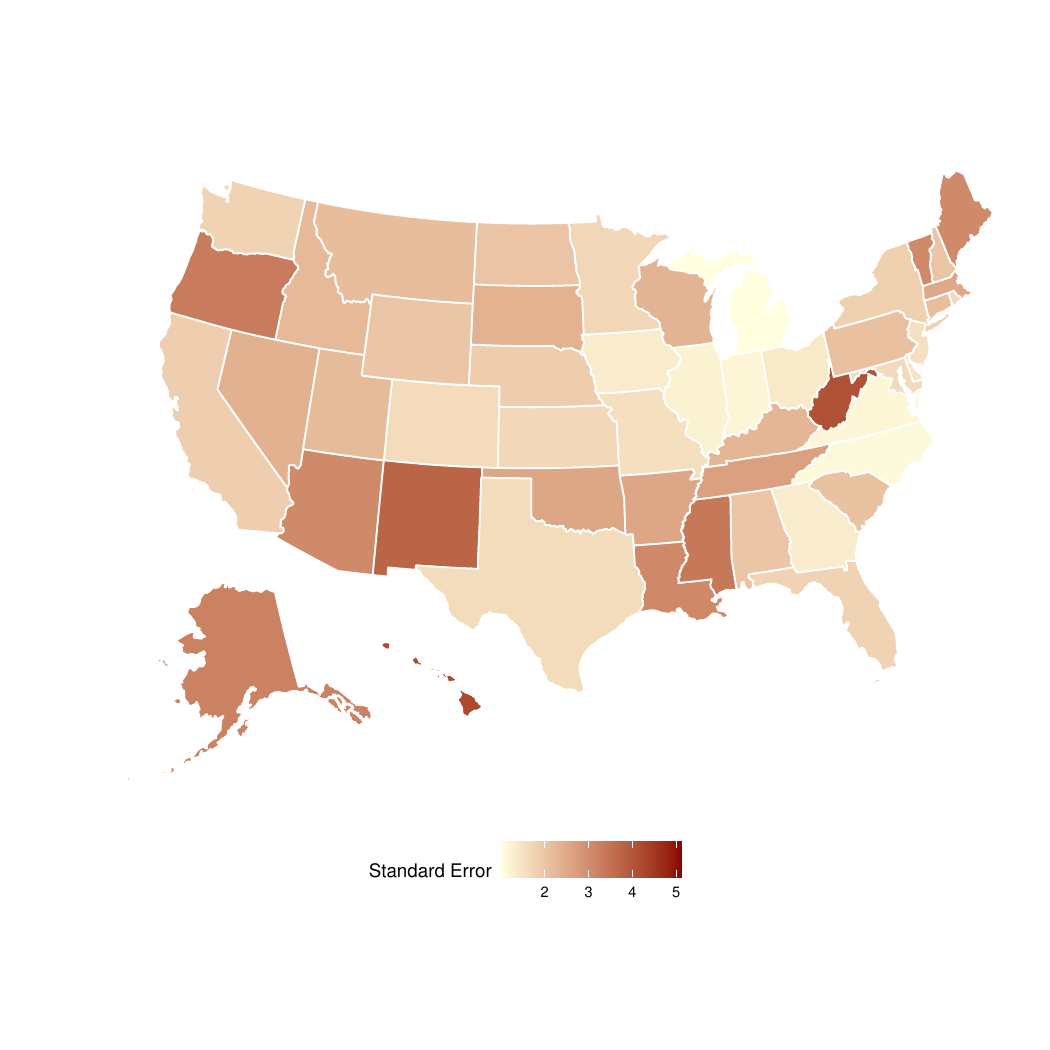}
        \caption{Standard Error estimates}
        \label{fig:se_map_plot}
    \end{subfigure}
    \caption{Statewise poverty rate estimates and standard errors obtained from the CBP model}
    \label{fig:benchmark_comparison}
\end{figure}
Figure~\ref{fig:estimate_map_plot} and Figure~\ref{fig:se_map_plot} show the benchmarked estimates of poverty rate and the corresponding standard errors from the proposed CBP model.
\begin{figure}[h]
    \centering
    \begin{subfigure}[t]{0.48\linewidth}
        \centering
        \includegraphics[width=\linewidth]{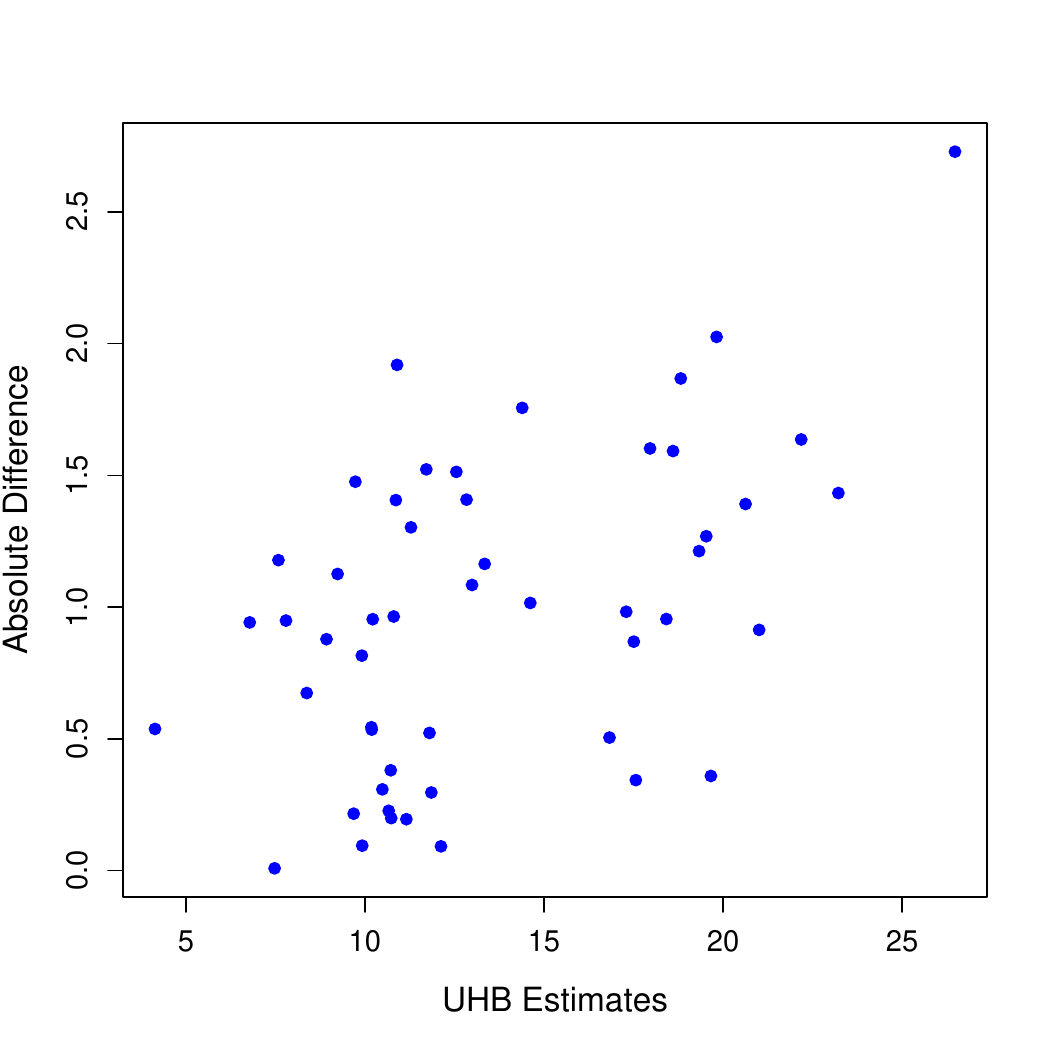}
        \caption{Absolute Difference between proposed benchmarked and standard estimates}
        \label{fig:diff_plot}
    \end{subfigure}
    \begin{subfigure}[t]{0.48\linewidth}
        \centering
        \includegraphics[width=\linewidth]{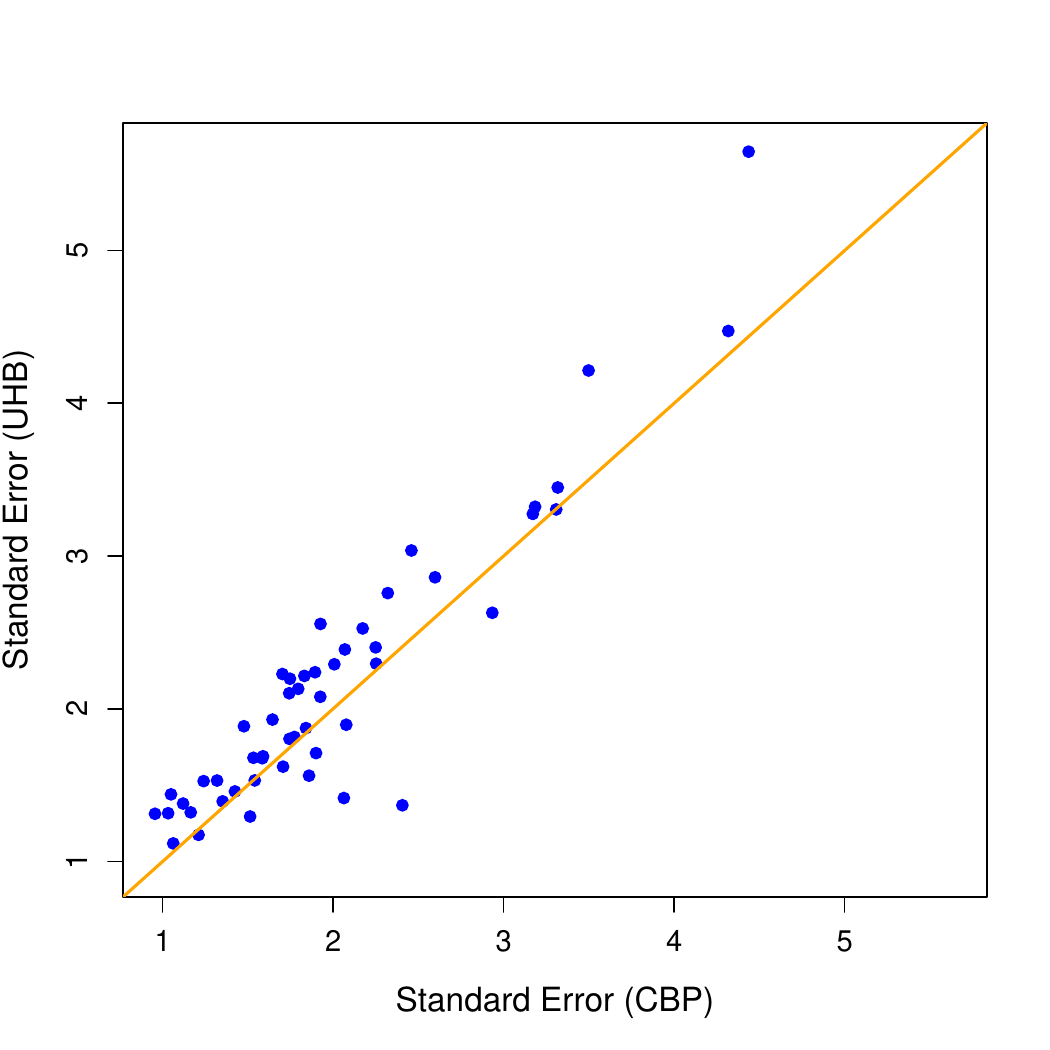}
        \caption{Comparison of model-based standard errors}
        \label{fig:se_plot}
    \end{subfigure}
    \caption{Comparison of results from the proposed benchmarked and UHB models across states.}
    \label{fig:benchmark_comparison}
\end{figure}
The aggregate of CBP estimates, aggregated with appropriate weights across the states, exactly equals to $15.9$\footnote{https://fred.stlouisfed.org/data/PP5T17US00000A156NCEN}, the national-level direct estimate. However, the similar aggregate of the standard Gaussian FH model estimates is $14.76$, indicating about 7.16\% deviation from the national level estimate. Figure~\ref{fig:diff_plot} plots the absolute difference between the CBP and UHB estimates against the UHB estimates. The differences are  all positive and tends to increase with the UHB estimate, indicating that benchmarking shifts the estimates more upwards in states with higher poverty rates. The model-based standard error for the benchmarked FH estimator is less compared to the standard FH estimator in most cases, and is shown in  Figure~\ref{fig:se_plot}.

\section{Discussion}
We have proposed a novel method to enforce exact benchmarking on the basic area-level FH model. The proposed benchmarked FH estimator, constructed through benchmarking prior construction, is comparable to the standard FH estimator in terms of estimation accuracy; in fact, it gives lower posterior standard deviations for most areas when the model is complex.  This approach does not require any post-hoc adjustments, making it a straightforward yet effective framework for constructing a benchmarked version of the FH estimator. Unlike post hoc adjustment-based methods, the proposed framework also accommodates an explicit prior on the coefficient vector $\bbeta$, allowing both the area level mean $\btheta$ and $\bbeta$ to be updated jointly, thus producing an explicit posterior for $ \bbeta$. Moreover, the availability of closed-form full conditional distributions for the benchmarked model facilitates efficient posterior sampling using Gibbs Sampling.
\par
Another key advantage is that, in the presence of data from nested geographical hierarchies (e.g., county-to-state-to-national aggregation), the proposed benchmarking prior construction can be extended to ensure exact benchmarking consistency at every level of aggregation while propagating the uncertainty across levels. This can be achieved by embedding the benchmarking priors for each level within a single joint Gibbs sampler, by drawing the higher-level (e.g., state-level) benchmarked estimates first and constructing the lower-level (e.g., county-level) benchmarking prior conditioning on the obtained higher-level posterior draws, allowing uncertainty estimates from the higher-level benchmarking steps to propagate naturally into the credible interval of the lower-level means.
\par
The benchmarking prior construction is applicable directly to variants of the Fay--Herriot models. We construct benchmarked versions of the robust Fay--Herriot model with multivariate-$t$ random effects and of the nonlinear Fay--Herriot model of \citet{parker2024nonlinear}, in which the mean function is represented by a random weight neural network. The implementation details of the latter are available in the Supplementary Material. In both cases, the max-fidelity prior preserves conjugacy; hence the Gibbs sampler retains closed-form full conditionals. For the robust model, the simulation results in the Supplementary Material show a similar pattern as compared to those for the Gaussian model. The CBP performs comparably to UHB under correct specification and yields clear gains in MSE and interval score when relevant covariates are omitted, while satisfying the benchmark exactly.

\vspace{-0.5em}
\section{Disclaimer}
This report is released to inform interested parties of ongoing research and to encourage discussion of work in progress. The views expressed are those of the authors and not those of the U.S. Census Bureau.

\bibliographystyle{abbrvnat}
\bibliography{references}


\end{document}